\documentclass[11pt]{article}

\usepackage{authblk} % have affiliations with article
\usepackage{graphicx}% Include figure files
\usepackage{epsfig}
\usepackage{dcolumn}% Align table columns on decimal point
\usepackage{bm}% bold math
\usepackage{amsmath}
\usepackage{amssymb}

\newcommand{\beq}{\begin{equation}} \newcommand{\eeq}{\end{equation}}
\newcommand{\bea}{\begin{eqnarray}} \newcommand{\eea}{\end{eqnarray}}

  \newcommand
{\Romannumeral}[1]{\uppercase\expandafter{\romannumeral#1}}

\newcommand{\be}{\begin{enumerate}} \newcommand{\ee}{\end{enumerate}}
\newcommand{\bi}{\begin{itemize}} \newcommand{\ei}{\end{itemize}}
\newcommand{\ba}{\begin{array}} \newcommand{\ea}{\end{array}}
\newcommand{\bc}{\begin{center}} \newcommand{\ec}{\end{center}}
\newcommand{\bt}{\begin{tabular}} \newcommand{\et}{\end{tabular}}

\def\lsim{\mathrel{\rlap{\lower4pt\hbox{\hskip1pt$\sim$}}
    \raise1pt\hbox{$<$}}}           % less than or approx. symbol
\def\gsim{\mathrel{\rlap{\lower4pt\hbox{\hskip1pt$\sim$}}
    \raise1pt\hbox{$>$}}}           % greater than or approx. symbol

\newcommand{\half}{\textstyle {1\over2} \displaystyle}    % One half
\newcommand{\thrqu}{\textstyle {3\over4} \displaystyle}   % Three quarters
\newcommand{\thrha}{\textstyle {3\over2} \displaystyle}   % Three halves
\newcommand{\Dslash}{{\hbox{D}\kern-0.6em\raise0.15ex\hbox{/}}} % D slash
\newcommand{\Dslashb}{{\hbox{{\bf D}}\kern-0.6em\raise0.15ex\hbox{/}}}% bold D slash

\renewcommand{\et}{\eta}

\begin{document}

\setlength{\oddsidemargin}{0cm} \setlength{\baselineskip}{7mm}

\begin{normalsize}\begin{flushright}

% CERN-PH-TH/2006-147 \\
% DAMTP-2006-59 \\
% UCI-2004-xx \\
October 2012 \\

\end{flushright}\end{normalsize}

\begin{center}
  
\vspace{5pt}

{\Large \bf Composite Leptons at the LHC }

\vspace{30pt}

{\sl Herbert W. Hamber}
$^{}$\footnote{e-mail address : Herbert.Hamber@aei.mpg.de} 
\\
Max-Planck-Institut f\"ur Gravitationsphysik \\
(Albert-Einstein-Institut)\\
Am M\"uhlenberg 1 \\
D-14476 Potsdam, Germany \\

\vspace{5pt}

and 

\vspace{5pt}

{\sl Reiko Toriumi}
$^{}$\footnote{e-mail address : RToriumi@uci.edu} \\
Department of Physics and Astronomy, \\
University of California, \\
Irvine, California 92697-4575, USA \\

\vspace{10pt}

\end{center}

\begin{center} {\bf ABSTRACT } \end{center}

\noindent

In some models of electro-weak interactions the W and Z bosons are
considered composites, made up of spin-one-half subconstituents.
In these models a spin zero counterpart of the W and Z boson naturally 
appears, whose higher mass can be attributed to a particular type of hyperfine
spin interaction among the various subconstituents.
Recently it has been argued that the scalar state could be
identified with the newly discovered Higgs (H) candidate.
Here we use the known spin splitting between the W/Z and H states
to infer, within the framework of a purely phenomenological model,
the relative strength of the spin-spin interactions.
The results are then applied to the lepton sector, and used to
crudely estimate the relevant spin splitting between the two lowest 
states.
Our calculations in many ways parallels what is done in the 
$SU(6)$ quark model, where most of the spin splittings between
the lowest lying baryon and meson states are reasonably well
accounted for by a simple color hyperfine interaction, 
with constituent (color-dressed) quark masses.

% \vspace{15pt}

% \begin{center} {\it (Submitted to the Physical Review D)} \end{center}
% \noindent PACS Numbers: 04.60.-m, 04.60.Gw, 04.60.Nc, 98.80.Qc
% Above are: Quant Grav, Cov Sum over Hist Quant, Latt Grav, Quant Cosm 

\vfill

%% no page number on 1-st page
\pagestyle{empty}

\newpage

\pagestyle{plain}

\section{Introduction}
\label{sec:intro}

\vskip 20pt

The idea that weak vector bosons and perhaps even leptons and quarks
are composite is not new [1-11].
Recently such composite models have been given new impetus by the
experimental discovery of a new Higgs-like particle \cite{atl,cms} with a 
mass significantly above the known weak vector bosons.
In \cite{fr12} it was suggested that the new spin-zero state
could be identified as the spin partner of the $W$ and $Z$ 
bosons, whose higher mass could arise because of the non-trivial
nonperturbative dynamics of the postulated subconstituents.

The question then arises of how such a scenario could be tested in practice,
and what notable observable differences would arise when compared
to the Standard Model.
One obvious consequence of the existence of subconstituents
is the deviation from point-like behavior at sufficiently high
energies and the
appearance of form factors, just as in the case of
the $\rho$ meson and most other hadrons composed of quarks.
Nevertheless the experimental measurement of form factors
is notoriously time-consuming and requires both high energies
(to probe in detail the inner structure) and high statistics.

Another option is to pursue the analogy with the quark model,
and derive a number of predictions that can be obtained in
a rather straightforward way from the nature of the 
constituents and their color, flavor and spin wave functions.
While these approaches have enjoyed some degree of success
for $QCD$ (summarized below), one additional obstacle
in the case of composite weak vector bosons is represented
by the fact that virtually nothing is known about the 
nonperturbative ground state of confining chiral gauge
theories.
The main reasons being that it is not easy to put
chiral fermions on the lattice, and also the rather serious issue
that the fermion determinant is generally complex
due to the anomaly.
In one of the simplest contexts the problem arises
because the fermion determinant
\beq
\det i \Dslashb_L = \left [ \det ( - \Dslashb_L^2 + \delta m^2 )
  \right ]^{1/2}
\eeq
is not real, since the operator
\beq
- \Dslashb_L^2 \; = \; - D_L^2  + \half \, g \, \sigma_{\mu\nu} \, F^{\mu\nu}
\eeq
is not Hermitean.
Here $\sigma_{\mu\nu} = { 1 \over 2 i } [ \sigma_\mu , \sigma_\nu ] $
with $\sigma_\mu = (1,{\bf \tau})$ and
$ - D_L^2 \; = \; - {\bf D}^2 
+ i \, D_0 \; {\bf \tau} \cdot {\bf D} $;
the second order formulation is used
to avoid the notorious lattice fermion doubling problem.
Since chiral symmetry is no longer explicit, the $\delta m^2 $
counterterm needs to be fine-tuned so as to obtain a purely 
left-handed fermion.
Also, while gauge invariance can be maintained in the second
order lattice formulation, the same might not be true
for Lorentz invariance, which should hopefully nevertheless be
recovered in the continuum limit. 
For an early perspective on these and other subtle
issues associated with the nonperturbative formulation
of chiral gauge theories see, for example, \cite{mate}.

\vskip 40pt

\section{Hyperfine Interaction}
\label{sec:wdweq}

We now turn to the formulation of a simple model
used later to account for the spin splittings
of the observed bound states.
In $QED$ for two charged particles with magnetic moments 
${\bf \mu}_i = (e_i / 2 m_i ) {\bf \sigma}_i $ 
and charges $e_i e_j \equiv 4 \pi \alpha $ one has
\beq
\Delta E \; = \; { 8 \pi \over 3 } \, { \alpha \over m_i m_j } \;
\vert \psi (0) \vert^2 \; {\bf S}_i \cdot {\bf S_j} \; .
\eeq
This applies to $s$-wave states only, for which
$\vert \psi (0) \vert^2 \neq 0 $.
For the ground state of Hydrogen one has for the wave function 
at the origin
\beq
\vert \psi_{100} (0) \vert^2 \; = \; { 1 \over \pi a_0^3 } 
\eeq
for a Bohr radius $a_0 \equiv \hbar^2 / m \, e^2 $.
Thus for the non-relativistic Hydrogenic case
$ \vert \psi_{100} (0) \vert^2 \; \simeq \; m^3 $ where
$m$ is the reduced mass. 
For the relativistic Dirac equation case the relevant 
results were obtained in \cite{bs57}.

Similar formulas can be written down for the case of $QCD$.
In the case of the color magnetic interaction of quarks
one has instead, from single gluon exchange,
\beq
\Delta E (q \bar q ) \; = \; { 32 \pi \over 9 } \, 
{ \alpha_S \over m_i m_j } \;
\vert \psi (0) \vert^2 \; {\bf S}_i \cdot {\bf S_j} \; ,
\eeq
for $q \bar q$ pairs in a color singlet meson, and
\beq
\Delta E (q q ) \; = \; { 16 \pi \over 9 } \, 
{ \alpha_S \over m_i m_j } \; 
\vert \psi (0) \vert^2 \; {\bf S}_i \cdot {\bf S_j} \; ,
\eeq
for $ q q $ pairs in a color singlet baryon.
In either case the coupling is proportional to $\alpha_S$,
with the different numerical coefficients due to
color factors as they apply to the relevant color
group representation \cite{clo}.
For the above formula to be useful, the parameters
$m_i$ are taken to be the dressed (or constituent) quark masses,
the latter reflecting the dressing of the current algebra
bare quark states by a large virtual gluon cloud.
The bound state wave function at the origin can be estimated
crudely in a non-relativistic potential model, but
in the end it would require a complex nonperturbative
(and relativistic) calculation,
and for practical purposes it is taken instead as a free 
parameter.

For two spins with total spin $ {\bf S} = {\bf s}_i + {\bf s}_j $ 
one has in general
\beq
{\bf s}_i \cdot {\bf s}_j \; = \; 
\half \left [ S(S+1) - s_i (s_i +1) - s_j (s_j +1) \right ]
\; = \; 
\begin{cases}
 - {3 \over 4}  & {\rm for } \; S=0
\\ 
+ { 1 \over 4 } & {\rm for} \; S=1 
\end{cases} \; .
\eeq
This last result can then be used to estimate
the spin splittings for mesons \cite{dgg,per}.
The wave function at the origin is then related to 
hadronic quantities such as $f_\rho$, the $\rho$ meson decay
constant.
One important aspects of vector-like $SU(N)$ gauge theories
is that chiral symmetry is dynamically broken.
Instead of the free field (or perturbative) result
\beq
< \bar \psi \psi > \; \simeq \; m_q^3
\eeq
with $m_q$ the (current algebra) light quark mass,
one instead has the nonperturbative $QCD$ result
\beq
< \bar \psi \psi > \; \simeq \; \Lambda_{\overline{MS}}^3
\eeq
which shows the non-vanishing nature of the $QCD$ fermion
condensate in the chiral limit $m \rightarrow 0$ \cite{hp81}.
In more practical terms, it is known that the condensate has
the value 
$ < \bar \psi \psi > \, \approx \, ( 300 \, MeV)^3 $, either
inferred from a direct lattice calculation, or from
the standard $PCAC$ relations involving the pion and 
current algebra quark masses.

In the case of three spins, where now
$ {\bf S} = {\bf s}_i + {\bf s}_j + {\bf s}_k $, one has instead
\beq
\sum {\bf s}_i \cdot {\bf s}_j \; = \; 
\half \left [ S(S+1) - 3 s(s+1) \right ]
\; = \; 
\begin{cases}
 - {3 \over 4}  & {\rm for } \; S= {1 \over 2}
\\ 
+ { 3 \over 4 } & {\rm for} \; S= { 3 \over 2}
\end{cases} \; .
\eeq
The latter is of course useful for the case of baryons, where
it can used both for the baryon octet ($s=\half$) and
decuplet $(s=\thrha)$.

In practice it is known that both sets of formulas give a reasonably 
good description of the (non-singlet) $S=0$ (pseudoscalar) and
$S=1$ (vector) meson multiplets, and
an equally reasonable description of the baryon
octet ($s=1/2$) and decuplet ($s=3/2$), with
some slight modifications to account for
the fact that for some baryons, like the $\Lambda$,
the large quark mass difference ($u/d$ vs. $s$) plays a
significant role \cite{dgg}.
In fact, the above formulae work much better than expected.
They reproduce the known light meson spectrum to
within a few percent, and the light baryon spectrum
(octet and decuplet) to a percentage or better.
The only exception of course are the isoscalar mesons
($\eta, \eta'$) which are affected by a large mixing with light
pseudoscalar glueballs (the so-called $U(1)$ QCD axial anomaly problem),
and therefore require additional input in
the form of suitable flavor mixing matrices.

\vskip 40pt

\section{Composite $W^\pm $ and $Z^0$ Bosons }
\label{sec:wboson}

In a number of composite vector boson scenarios the
$W$ and $Z$ weak vector bosons are made up of spin-$\half$
subconstituents \cite{fr12}.
In these theories it is generally assumed that the leptons,
and perhaps even the quarks, are composite.
Here we will consider, for simplicity, what we regard as one of the simplest
scenarios, where the leptons are composed of three
spin-$\half$ elementary constituents, possibly not even 
exactly of the same mass ($h h h$, and ${\bar h}{\bar h}{\bar h}$
for their antiparticles).
Furthermore, we will assume that the $W$ and $Z$ bosons are made of 
six of the same elementary constituents,
three of them particles and the rest antiparticles
($h h h {\bar h}{\bar h}{\bar h}$).
\footnote{
Here we are mainly concerned with the spin splittings arising
from some hypothetical confining chiral hypercolor force.
The relevant electric charge assignments for the subconstituents
were given explicitly in \cite{har}. Thus the $W^+$ boson is
made up of six $TTTVVV$ spin-$\half$ constituents, while the $Z^0$ contains
both ${\bar T}{\bar T}{\bar T} TTT $ and 
$ {\bar V}{\bar V}{\bar V} VVV$, with $V$ having charge $q=0$ and
$T$ charge $q=1/3$.
In this model the electron is ${\bar T}{\bar T}{\bar T}$ and
the electron neutrino $VVV$.
The heavier second and third generation leptons are then accounted 
for by the presence of more massive partners to the $T$ and $V$, 
or perhaps by some (unknown) dynamics leading to radial excitations.}
Later on we will also look, for completeness, at the case of
$W$'s and $Z$'s made out of ($h h {\bar h}{\bar h}$)
or even ($h {\bar h} $).

For concreteness, and in a first approximation, we will assume that all
subconstituents have roughly the same mass, and
that the binding in hypercolor singlet states occurs due to a 
hypercolor force based on the group $SU(N)$.
Not much is known about the nonperturbative
ground state of chiral gauge theories, and
therefore about the nature of the effective hypercolor spin
interactions, or about the hypercolor dressing of bare
(current algebra) subconstituents.
In the following we will pursue the simplest assumptions
in order to obtain a series of admittedly rather crude estimates
for the effective spin dependent forces.
A more sophisticated estimate for the bound state
energies is not possible yet, given our almost complete
ignorance regarding the properties of the hypercolor 
symmetry group, the nature of the subconstituents, and
the general nonperturbative properties
of chiral gauge theories.

In the following the spin interaction will be modeled
after the $QCD$ one described earlier.
For $N$ spin-$\half$ objects, where now 
$ {\bf S} = \sum_i {\bf s}_i $, one obtains
\beq
\sum_{i>j} {\bf s}_i \cdot {\bf s}_j \; = \; 
\half \left [ S(S+1) - N \cdot \thrqu \right ]
\eeq
Thus each ``bond'' contributes the same amount
\beq
S_{12} \; \equiv \; < {\bf s}_i \cdot {\bf s}_j > \; = \; { 1 \over N (N-1) }
\left [ S(S+1) - N \cdot \thrqu \right ]
\eeq
Thus for six spin-$\half$ objects one has
\beq
< \sum {\bf s}_i \cdot {\bf s}_j > \; = \; 
{ 1 \over 30 } \left [ S(S+1) - 6 \cdot \thrqu \right ]
\; = \; 
\begin{cases}
 - {3 \over 20}  & {\rm for } \; S= 0
\\ 
- {1 \over 12} & {\rm for} \; S= 1
\end{cases} \; .
\eeq
A first-principle estimate of the spin interactions
would seem to require a nonperturbative calculation
in a confining chiral gauge theory.
But, as alluded to previously, the nature of the effective
spin interaction between subconstituents in a 
nonperturbatively treated chiral gauge theory is not well understood.
Here the assumption will be made that it
can be written in the form of an effective spin-spin interaction of
the type used for hadron models, with two unknown
parameters characterizing the strength of the 
fermion-anti-fermion ($\alpha$) and fermion-fermion
($\beta$) direct spin interactions.

In the following we will only consider the $l=0$ $s$-wave states.
Then, if indeed the $S=0$ and $S=1$ bosons are made out
of six subconstituents ($h h h {\bar h}{\bar h}{\bar h}$, 
three fermions and three anti-fermions), one can write
\beq
M(m,S) \; = \; 6 \, m \; + \, 
{ 1 \over m^2 } \left ( \, 9 \, \alpha \, S_{12} + 6 \, \beta \,
S_{12} \, \right )
\eeq
with numerical coefficients reflecting the number
of fermion-fermion (6) and fermion-anti-fermion (9) 
interaction bonds.
Here the parameter $m$ stands for the dressed  mass of the 
subconstituents, arising mainly from the strong hypercolor 
confining gauge interactions.
The spin interaction is modeled, again, after $QED$
and single-gluon exchange $QCD$, and
$\alpha$ and $\beta$ are taken so far as entirely free real parameters.
In $QED$ one has of course $\alpha / \beta = -2 $.

For the spin zero weak boson ($H$) one obtains
\beq
M(m,0) \; = \; 6 \, m - { 27 \, \alpha \over 20 \, m^2 } - 
{ 9 \, \beta \over 10 \, m^2 }
\eeq
and for the spin one weak boson ($W$)
\beq
M(m,0) \; = \; 6 \, m - { 3 \, \alpha \over 4 \, m^2 } - 
{ \beta \over 2 \, m^2 } \; .
\eeq
The mass splitting is therefore given by
\beq
M(S=0) \, - \, M(S=1) \; = \; 
{1 \over 5 \, m^2 } \left ( -3 \, \alpha - 2 \, \beta \right ) \; .
\eeq
The known mass of the $S=1$ $W^\pm$ boson ($m_W = 80.385 \, GeV$)
and of the $S=0$ Higgs candidate ($m_H = 125.9 \, GeV$) then
fixes two of the free parameters, say $\alpha$ and the 
constituent fermion mass
\beq
m \; = \; 3.915 \, GeV \; .
\eeq
Note that we used the $W$ boson mass instead of the $Z$ mass (or some
combination of the two), since we took into consideration the fact
that the $Z$ mass is shifted upward by the mixing with the photon,
which does not however affect the $W$ mass.
In addition one finds $\alpha / m^3 = -22.04 $.

An alternate possibility, advocated explicitly in \cite{fr12},
is to consider the $W$ and $Z$ bosons as made exclusively of
two spin-$\half$ subconstituents, without any further ingredients
($ h {\bar h}$).
If that is indeed the case, then one has simply for the 
fermion-antifermion pair
\beq
{\bf s}_i \cdot {\bf s}_j \; = \; 
\begin{cases}
 - {3 \over 4}  & {\rm for } \; S=0
\\ 
+ { 1 \over 4 } & {\rm for} \; S=1 
\end{cases} \; ,
\eeq
and therefore
\beq
M(m,S) \; = \; 2 \, m \, + \, {\alpha \over m^2} \, S_{12} \; .
\eeq
From the known $W$ and $Z$ masses one then obtains,
not unexpectedly, a much larger value for the subconstituent 
effective mass $m \approx 45.88 \, GeV $,
as well as a spin coupling of magnitude $\alpha / m^3 = -0.992 $.

\vskip 40pt

\section{Composite Leptons}
\label{sec:leptons}

To fix the value of the remaining spin interaction parameter 
$\beta$ one turns next to the leptons.
For concreteness we focus here on
the electron, assumed to be made out
of three spin-$\half$ subconstituents.
As before, we will consider only the $l=0$ $s$-wave states.
Then, in analogy to the quark model baryon octet and decuplet,
the corresponding lepton state wave function would be constructed in
accordance with the hypercolor analog of the quark model 
non-relativistic $SU(6)$ flavor-spin symmetry.
Nevertheless, here we will not need to make use of any specific
details of the subconstituent wave functions, and rely instead
only on the spin content of the lepton subconstituents.

Then for the lepton case one has simply
\beq
\mu(m,S) \; = \; 3 \, m + { 1 \over m^2 } \cdot 3 \beta \, S_{12} 
\eeq
with here
\beq
S_{12} \; = \;  { 1 \over 6 }
\left [ \, S(S+1) - 3 \cdot \thrqu \, \right ]
\; = \; 
\begin{cases}
 - {1 \over 4}  & {\rm for } \; S= { 1 \over 2}
\\ 
+ {1 \over 4} & {\rm for} \; S= { 3 \over 2}
\end{cases} \; .
\eeq
After using the known value for the electron mass 
$ m_e = 0.511 \, MeV $, one is finally
able to fix the parameter $\beta$ as well.
One finds
\beq
{ \alpha \over m^3 } \; = \; - 22.04
\eeq
\beq
{ \beta \over m^3 } \; = \; 4.00
\eeq
and thus for the ratio
\beq
{ \alpha \over \beta } \; = \; - 5.51 \; .
\eeq
For the spin-$\thrha$ heavy lepton one obtains the estimate
\beq
\mu_{3/2} \; = \; 23.49 \, GeV \; .
\eeq
Note that if one had given a zero mass to the lightest
$s= \half $ lepton (like a neutrino) then
the above numbers would have hardly changed (for this case 
one has exactly $\beta / m^3 = 4 $), and one would
have still obtained $ \mu_{3/2} \approx 23.49 \, GeV $.
The main reason for this lies in the fact that the electron is,
for practical purposes, already very {\it light}.
\footnote{
As an exercise, one can explore how the mass of the
spin-$\thrha$ heavy lepton and the parameter $\beta$ are
shifted when one uses instead the mass of the muon 
(for $m_\mu = 105.658 \, MeV$ one finds $\mu_{3/2} = 23.39 \, GeV$ 
and $\beta / m^3 = 3.968 $)
or the mass of the tau 
(for $m_\tau = 1776.82 \, MeV$ one finds $\mu_{3/2} = 21.71 \, GeV$ 
and $\beta / m^3 = 3.395 $). 
So there is some change, but it is not too large.}
One potential problem with this rather low mass excited lepton 
is that it might already be largely 
excluded by the $LEP$ data on $Z^0$ decays \cite{hl10}, see
discussion below.

Again one can revisit here the alternate possibility of
\cite{fr12} where the 
the $W$ and $Z$ bosons as made exclusively of
two spin-$\half$ subconstituents ($h {\bar h}$).
Then the leptons and quarks contain a single fermion
and also an additional hypercolor-carrying
scalar (spin-zero) constituent, of unknown mass.
Generally elementary scalars suffer from quadratic
mass divergences, so it is unclear how such a
state could generate the relatively light masses
of quarks and leptons.
To make further progress, we will assume here
that such a scalar state is itself a bound state,
or condensate, of either two fermions 
($h h$) or a fermion-antifermion pair ($h {\bar h}$).
Then, if the spin interactions between the three
subconstituents is treated on an equal basis,
one has for the first case $\beta / m^3 \approx 4 $
and in the second case $\beta / m^3 \approx 14 $,
where $m$ and $\alpha$ are fixed, as before, by the $W$ mass.
In either case though one finds the same result
for the spin-$\thrha$ heavy lepton, namely
$ \mu_{3/2} \approx 275.3 \, GeV $.
This last case would suggest
that the more subconstituents
are arranged into the $W$ and $Z$ bosons, the lighter
the heavy spin-$\thrha$ lepton can be made.
So, one way of viewing our (admittedly very simple) results is that
if no light s=$\thrha$ lepton (we find about $23.49 \, GeV$) is observed,
that would exclude the case of a $W$ or $Z$ made out of many (six) 
subconstituents, and favor instead a scenario where fewer 
(say two) subconstituents make up the weak vector bosons.
For completeness we will also quote here the excited lepton
mass obtained if one assumes that the $W$ and $Z$ bosons are 
made of four spin-$\half$ subconstituents ($h h {\bar h} {\bar h}$).
These bound states are similar to the four-quark exotic states
of $QCD$ considered in \cite{ma04}.
Then one has $ \mu_{3/2} \approx 86.44 \, GeV $, which
lies between the two previous cases.
Again it would seem that this value could already be excluded
by the $LEP$ data on $Z^0$ decays.

Let us note here that in the quark model the 
main source of uncertainty in predicting
the splittings in the baryon multiplets from the
meson octet, or vice versa, is the fact that the wave function
at the origin is not quite the same for the two
type of states; 
in fact it is known that then r.m.s. charge radius of mesons
($r_0 \approx 0.6 \, fm$) is smaller than the corresponding
quantity for baryons  ($r_0 \approx 0.8 \, fm$),
which when cubed gives rise to roughly a factor of two
correction, and in the right direction to account for the
observed hadron spin splittings.
If that is the case here as well (in the sense that the r.m.s.
charge radius for the electron could be 
significantly smaller than the corresponding quantity
for the vector and scalar weak bosons, 
when taking into account
the fact that the scalar and vector bosons contain twice as 
many subconstituents compared to the leptons), 
then this would suggest that
the parameter $\beta$ varies a bit depending
on which states are considered (composite leptons vs.
composite vector and scalar bosons).
Nevertheless, the subconstituent mass $m$ is expected to stay
the same, and $\beta$ is after all determined here 
only from the known electron mass, so the spin splitting
in the lepton sector remains unchanged, at least in
the context of this rather simple model.

From a non-relativistic perspective (which might very well be
totally inadequate in the present context) one has 
$\alpha, \beta \simeq \vert \psi (0) \vert^2 $, giving therefore
in light of the previous results
$\vert \psi (0) \vert^2 \simeq m^3 $.
For quark-antiquark pairs the latter wave function value
is related, via the 
Van Royen-Weisskopf formula and its $QCD$ extensions \cite{bar},
to the leptonic decay width of a vector boson,
\beq
\Gamma ( \, V \rightarrow l \bar l \, ) \; \approx \;
{ 16 \pi \, \alpha^2 \, e_Q^2 \over M_V^2 } \; 
\vert \psi (0) \vert^2 \; \left [ \, 1 + {\cal O} \left ( \alpha_S
  \right ) \, \right ]
\eeq
where $\alpha$ here is the usual $QED$ fine structure constant, 
$\alpha_S$ the strong $QCD$ coupling, and
$e_Q$ the relevant quark charge in units of the proton
charge.
All one can say in the present context is perhaps
that in the above formula, translated to the weak
vector boson context, one would expect
as a rough order of magnitude estimate (and not much more)
$\vert \psi (0) \vert^2 \simeq m^3 $, where
$m$ is the subconstituent effective mass given previously.

As far as production processes are concerned \cite{atl1,che}, 
if this light spin-$\thrha$ lepton indeed exists,
then it should be eventually observed along the
standard decay modes for the $Z$, such as $Z \rightarrow e^+ e^- $,
with the new spin-$\thrha$ lepton replacing the standard electron.
\footnote{In the absence of a specific model describing the
subconstituent's interaction, one can make the following very
rough estimate for various decay rates.
In the case of the spin-$\thrha$ composite lepton $e^*$ all three spins of
the subconstituents are aligned, and the same will of
course be true individually when such a pair is produced in the decay
of a weak boson ($W$, $Z$ or $H$).
On the other hand within the weak bosons themselves, and therefore before
such a decay, the subconstituent's spins are mostly anti-aligned, 
to account for the comparatively low total spin (zero or one) of
the $W$, $Z$ or $H$.
Consequently the subconstituent spin re-arrangement which is required 
in, say, a decay $ Z \rightarrow e^{*+} e^{*-} $ is significant.
One can argue that this should lead to a rather significant suppression, by one
or two orders of magnitude or perhaps even more, of this last
decay rate versus the more standard $ Z \rightarrow e^{+} e^{-} $.
Note that in the latter case the subconstituent's spins are
largely anti-aligned both in the initial and final state, leading
to essentially no significant spin re-arrangement suppression.
The above arguments could therefore provide a hint as to
why excited leptons have not been seen so far in $Z$ or $W$
decay.}
The present models gives few prediction about the width
of the new composite state, but it could be quite broad (as in
the case of the $\rho$ meson for which 
$ \Gamma_\rho / m_\rho \approx 776\, MeV / 149\,  MeV \approx 5 \% $).
In any case a credible estimate for the branching ratios of the new
excited lepton would clearly require at some point an understanding 
of the structure
of the underlying interaction Hamiltonian, from which the relevant
matrix elements would then be computed.
It is well known that such branching ratio estimates are already
hard to obtain in QCD, where the effects of the underlying confining
gauge dynamics play an important role.
It is also clear, from the QCD analogy, that a simple estimate
of the spin splittings between hadrons (as used here), based on an 
effective and largely incomplete model of single gluon exchange, 
does not extend or translate in a simple way to an
estimate of hadronic transition rates, especially
for the lighter hadrons, where chiral symmetry and current algebra
arguments play an important role.

%\newpage

\vspace{20pt}

{\bf Acknowledgements}

The authors are very grateful to prof. Harald Fritzsch for 
discussions and correspondence.
The work of H.W.H. was supported in part by the Max 
Planck Gesellschaft zur F\" orderung der Wissenschaften, and
by the University of California.
He wishes to thank prof. Hermann Nicolai and the
Max Planck Institut f\" ur Gravitationsphysik (Albert-Einstein-Institut)
in Potsdam for warm hospitality. 
The work of R.T. was supported in part by a DED GAANN Student Fellowship.
She particularly wishes to thank prof. Daniel Whiteson 
(LHC Atlas collaboration) for discussions and references.
H.W.H acknowledges useful correspondence with Fabiola Gianotti of the 
CERN LHC Atlas collaboration, and Roberto Tenchini of the CERN LEP Aleph
collaboration.

\newpage

\vskip 40pt

% \newpage

\vfill

%\newpage

\end{document}